\documentclass[final,5p,times,twocolumn]{elsarticle}
\usepackage{color}
\usepackage{graphicx}% Include figure files
\usepackage{bm}% bold math
\usepackage{booktabs}
\usepackage{lineno}
\usepackage{subcaption}
\def\Journal#1#2#3#4{{#1}\, {\bf #2}, #3 (#4)}
  
\def\EPH{Eur.\,Phys.\,H.}
\def\EPJ{Eur.\,Phys.\,J.}
\def\EPJA{Eur.\,Phys.\,J. A}

\def\NIMA{Nucl.\ Instrum.\ Meth.\ A}
\def\NPA{Nucl.\ Phys.\ A}
\def\NPB{Nucl.\,Phys.\,B}
\def\PLB{Phys.\ Lett.\ B}
\def\PR{Phys.\ Rept.}

\def\PRD{Phys.\,Rev.\,D}
\def\PRC{Phys.\,Rev.\,C}
\def\RMP{Rev.\,Mod.\,Phys.}

\def\JHEP{JHEP}
\def\PPNP{Prog.\,Part.\,Nucl.\,Phys.}
\def\APP{Acta Phys.\ Polon.}

\begin{document}

\begin{frontmatter}

%%%\preprint{}

\title{Observation of Feynman scaling violations and evidence for a new resonance at RHIC}

\address[label1]{Brookhaven National Laboratory, Upton, New York 11973, USA}
\address[label2]{Christopher Newport University, Newport News, Virginia 23606, USA}
\address[label3]{University of California, Berkeley, California 94720, USA}
\address[label4]{Institute of High Energy Physics, Protvino 142281, Russia}
%\address[label5]{Thomas Jefferson National Accelerator Facility, Newport News, Virginia 23606, USA}
\address[label6]{Shandong University, Jinan, Shandong 250100, China}
%\address[label7]{University of Virginia, Charlottesville, Virginia 22903, USA}
\address[label9]{University of Zagreb, Zagreb, HR-10002, Croatia}
\address[label11]{Department of Theoretical Physics, University of the Basque Country UPV/EHU, 48080 Bilbao, Spain}
\address[label12]{IKERBASQUE, Basque Foundation for Science, 48011 Bilbao, Spain}
\address[label13]{Temple University, Philadelphia, Pennsylvania 19122, USA}
\address[label14]{Kent State University, Kent, Ohio  44242, USA}

\author[label1]{L.~C.~Bland}
\author[label2]{E.~J.~Brash}
\author[label3]{H.~J.~Crawford}
\author[label4]{A.A.~Derevschikov}
\author[label1]{K.~A.~Drees}
\author[label3]{J.~Engelage}
\author[label1]{C.~Folz}
\author[label3]{E.~G.~Judd}
\author[label6,label1]{X.~Li}
\author[label4]{N.~G.~Minaev}
\author[label2]{R.~N.~Munroe}
\author[label4]{L.~Nogach}
\author[label1]{A.~Ogawa}
\author[label3]{C.~Perkins}
\author[label9]{M.~Planinic}
\author[label13]{A.~Quintero}
\author[label11,label12]{G.~Schnell}
\author[label14]{P.~V.~Shanmuganathan}
\author[label9,label1]{G.~Simatovic}
\author[label13]{B.~Surrow}
\author[label1]{T.~G.~Throwe}
\author[label4]{A.~N.~Vasiliev}

\address{\rm\normalsize(A$_N$DY Collaboration)$^*$}
%\address{(A$_N$DY Collaboration)\corref{cor1}}
\cortext[cor1]{\it{URL: www.andy.bnl.gov}}

%%%\collaboration{A$_N$DY Collaboration}\homepage{www.andy.bnl.gov}\noaffiliation

\date{\today}

\begin{abstract}

We report measurements of forward jets and dijets produced in Cu+Au collisions at
$\sqrt{s_{NN}}=200$ GeV at the Relativistic
Heavy Ion Collider.  We also report dijet production cross sections in p+p
collisions at $\sqrt{s}=510$ GeV. We use the invariant
dijet mass to search for indications of new particles.  The p+p
dijet results are compatible with string fragmentation models tuned to fit
LHC data.  The Cu+Au jet results far exceed Feynman scaling limits, and
are compatible with models that incorporate string fusion to increase parton energy,
acting as a QCD accelerator.  The Cu+Au dijet results can be mostly explained
by double parton scattering due to a parton flux from multiple p+p interactions
with $\sqrt{s}>>\sqrt{s_{NN}}$.  Further indication of the increased
parton energy is obtained from evidence of single- and
double-$\Upsilon$(1S) production in the forward direction in Cu+Au
collisions.  Finally, we report evidence for the production of a new resonance,
reconstructed from its dijet decay.
  
\vspace{0.2cm}
%\noindent PACS numbers:~~~12.38.Qk,13.87-a,13.88+e
\end{abstract}

\begin{keyword}
  forward jet and dijet production, Feynman scaling, QCD accelerator, tetraquark
\end{keyword}
  
\end{frontmatter}

%\begin{linenumbers}

The charges of QCD are confined in color neutral objects such as mesons and
baryons.  Many predictions have been made for more complex color
neutral objects, and high energy colliders have
recently identified candidates for four-quark and five-quark
configurations built around heavy quarks \cite{ALS17,OSZ18}.  Given theoretical
expectations that relativistic heavy-ion collisions produce
particles by parton recombination \cite{Fr03}, it is of interest to see if
standard particle search techniques can identify new color neutral
objects produced in a relativistic heavy-ion collision. 

Forward particle production is characterized by the produced
particles having a significant fraction of the momentum of the
beam, as established by the Feynman-$x$ scaling variable
($x_F=2p_z/\sqrt{s}$, defined in the center of mass with $z$ along the
beams).  At large collision
$\sqrt{s}$, the hadrons resolve themselves to their partons, each
carrying a fraction $x_i$ of the momentum from the $i=1,2$ incident hadrons. 
Large $x_F$ corresponds to $x_1>>x_2$, where $i=1$ is the hadron
heading towards the produced particles.  Large $x_F$ also probes the
lowest $x_2$ at a given $\sqrt{s}$, and is therefore interesting from the
standpoint of low-$x$ physics. 
At low-$x$ and large $\sqrt{s}$, the gluon density in a hadron is
theoretically expected to saturate.  Gluon saturation has been
identified as being responsible for small particle multiplicities in
heavy-ion collisions at large $\sqrt{s}$, and is an expected doorway
to the formation of quark-gluon plasma \cite{LM06}.

Nearby to particles produced with large $x_F$ is a significant flux of
partons that are spectators to a hard scattering event.  The large
density of forward partons opens the prospect for recombination 
of these many partons for the production of exotic particles.  The expectation of intense gluon
fields and large fluxes of partons in the forward direction make it
interesting to search for the production of new particles in this acceptance.

We report the cross sections for forward jet pair production in p+p
collisions at $\sqrt{s}$=510 GeV and forward jet and dijet production in Cu+Au collisions at
$\sqrt{s_{NN}}=200$ GeV.  Jets in p+p collisions were identified using the anti-kT algorithm \cite{Ca08}.
Jets in Cu+Au collisions were identified using the anti-kT algorithm and 
results were verified
by independent analysis using the Fastjet 3.3.2 package \cite{FJ11}.
Measurements were completed with a forward
calorimeter wall that had good response to both incident
electromagnetic and hadronic particles that are produced by colliding beams.
The measurements were made at interaction point (IP) 2, at the
Relativistic Heavy Ion Collider (RHIC) at Brookhaven National Laboratory in
2012.

The apparatus was previously discussed in our report of forward jet
production in p+p collisions at $\sqrt{s}=510$ GeV \cite{Bl15}.  In
short, 236 cells were used to make a 200 cm $\times$ 120 cm forward
calorimeter wall, with a central $(20~$cm$)^2$ hole for the beams.
Each cell was $117~$cm $~\times~ (10~$cm$)^2$ of lead, with an embedded
matrix of $47\times47$ scintillating fibers that ran along the cell
length in a spaghetti calorimeter configuration \cite{Ar98}.  The
calorimeter wall was positioned 530 cm from the IP.  In addition, two
annular arrays of 16 scintillator tiles each were positioned at $\pm150$ cm to serve
as beam-beam counters (BBC) \cite{Bi01}.  The calorimeter had
$\sim$5.9 hadronic interaction lengths and $\sim$150 radiation lengths
of material, so was ideal for finding jets. The
calorimeter spanned the pseudorapidity range of about $2.4<\eta<4.5$ for
particles produced at the center of the vertex-z distribution and faced the Cu beam.  The
BBC reconstructed the $z$ component of the collision vertex from timing
measurements.  The BBC also is used to measure total charge, which
through simulation is related to the impact parameter of the colliding
ions.

The calibration of the calorimeter was previously described
\cite{Bl15}.  Peaks from minimum ionizing particles (MIP) from cosmic ray
muons were matched to set the hardware gain of each cell.  Software
relative gain corrections were made to match the slopes of the steeply
falling charge distributions for each cell from collision data.  The
absolute energy scale was determined from reconstruction of neutral
pions from pairs of photons detected in the calorimeter.  The
difference between the hadronic and electromagnetic response was
initially determined by simulation, and then confirmed by test-beam
measurements at FermiLab.  The calorimeter cell response to either
p+p or Cu+Au collision data can be described by full simulation.

The calibration of the BBC included gain matching based on the
MIP peak position and arrival time
matching of each BBC tile.  The MIP peak was adjusted so that its most
probable value was 100 ADC counts for each tile.  The individual tile
charge distributions are well represented by Cu+Au collision events
simulated by the HIJING event generator \cite{WG91}, subsequently run
through the GEANT~\cite{GEANT} model of the apparatus.  The calibrated charge sum
from the 16 detectors that comprise the BBC annulus that faces the Au
beam ($\Sigma Q_Y$) is used in the analysis of Cu+Au collisions to limit analyses to
semi-peripheral collisions.  The $\Sigma Q_Y$ is limited to 8000 counts,
which from HIJING/GEANT simulations corresponds to a minimum impact
parameter of 8 fm.  The jet results we report correspond to
grazing nuclear collisions to overlap of half of the Cu nucleus with
the Au nucleus.

The p+p data reported in this paper are from collisions at
$\sqrt{s}=510$ GeV, with 2.5 pb$^{-1}$ of integrated luminosity
accumulated, as measured using event rates calibrated in one dedicated vernier scan.  Both minimum bias (MB)
triggers and jet triggers were recorded.

The Cu+Au data reported here are from collisions at
$\sqrt{s_{NN}}=200$ GeV recorded later in the 2012 run using a MB
trigger, an inclusive jet trigger, and a dijet trigger.  The hardware
gains of the calorimeter were unchanged from p+p data.  The Cu+Au data were
obtained as a test of pulse-shape discrimination in the calorimeter.
Since the data were for an apparatus test, no vernier scan was
made to measure the colliding beam luminosity.  Consequently, we
report fraction of Cu+Au MB as a yield measure.  The MB trigger
requires hits in both BBC annuli.  The jet trigger sums ADC values
from the original modular calorimeters that were to the left and to
the right of the oncoming Cu beam excluding the outer two
perimeters of cells from these modules (see Ref.~\cite{Bl15}). 
The calorimeter was made annular by mounting cells above and below
the beams, closing the gap
between the original left and right modules,
although these cells were not used in the trigger.
The dijet trigger requires a coincidence between the left
and right jet patches.  The equivalent number of MB events from the
jet and dijet triggers were determined from emulations of those
triggers applied to MB events.

\begin{figure}[!htbp]
  \begin{tabular}{ll}
    \includegraphics[height=1.6in]{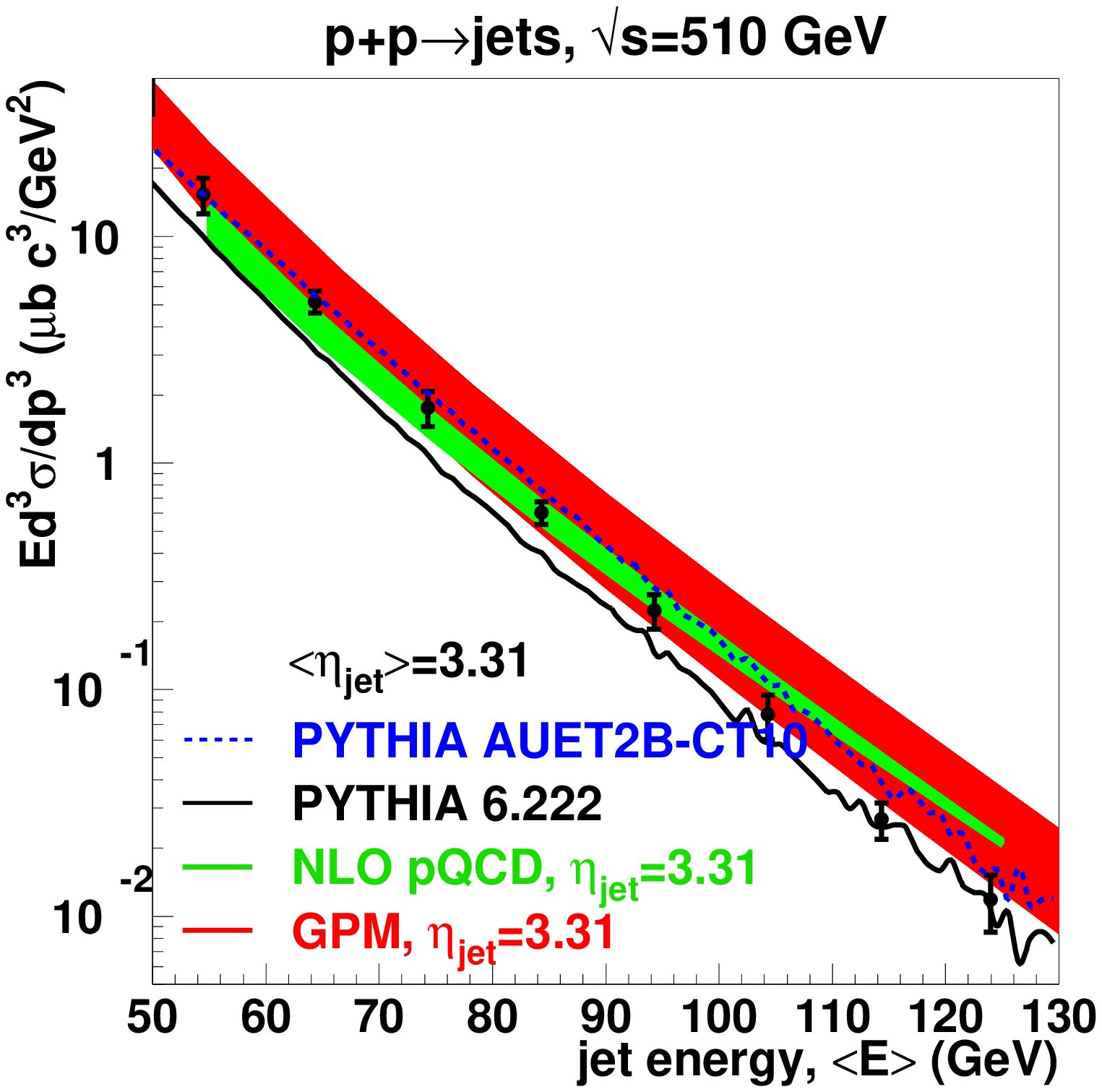} &
    \includegraphics[height=1.6in]{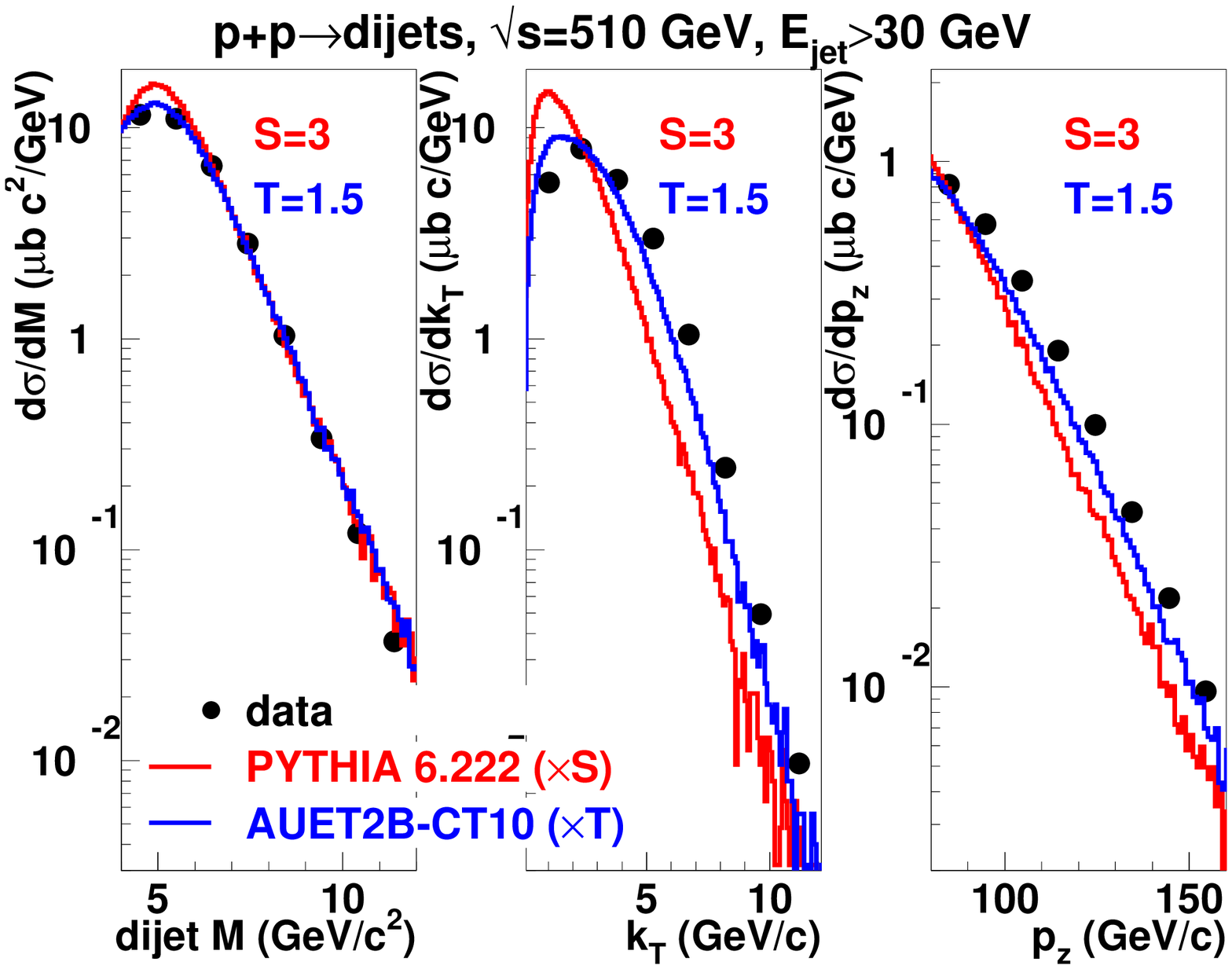}
  \end{tabular}
  \caption{(left) Inclusive forward jet production in p+p collisions
    compared to pQCD calculations (see Ref.~\cite{Bl15}) and string-model calculations, with
    AUET2B-CT10 representing a tune developed to explain LHC jet
    production.  (right) forward dijet cross sections from p+p
    collisions versus dijet mass, dijet transverse momentum,
    and dijet longitudinal momentum.  The combination of forward jet
    and forward dijet cross sections is very sensitive to string model
    tunings, and can be represented by tunes that explain LHC data,
    such as AUET2B-CT10, scaled by the factor $T$ in the plot.}
  \label{dijetcs}
\end{figure}

Following calibration, jets were reconstructed using the anti-$k_T$
algorithm \cite{Ca08} with a cone radius of $R_{jet}$=0.7 radians in
$(\eta,\phi)$ space. The modular jet finder used in \cite{Bl15} was
converted to an annular jet finder by proper treatment of the cyclic
$\phi$ variable.  The jet-energy-dependent efficiency of the jet trigger in p+p
collisions was determined from comparison of MB and jet
triggered data.  The jet reconstruction efficiency was determined from
full simulations that accurately describe the data, as further
discussed in Ref.~\cite{Bl15}.  For this
efficiency, particle jets, which are reconstructed from the simulation event
generators, are compared to tower jets, which are reconstructed from a GEANT
simulation of the apparatus in the same manner as is done for the
data.  The jet data in the left panel of Fig.~\ref{dijetcs} are from
Ref.~\cite{Bl15}.  It is included here for comparison to different
PYTHIA \cite{Sj06} tunes.

QCD processes that produce jets in hadroproduction always result in
multiple jets.  Results from the jet finder and the reconstruction of the
$z$ component of the collision vertex allow for attributing a four
momentum to each jet, assuming that the partons that give
rise to the jet are massless.  Inclusive pairing of valid jets is then
done, where a valid jet is within the calorimeter acceptance
($3.0<\eta<3.5$, stated elsewhere in the text as
$|\eta-\eta_0|<d\eta$, with $\eta_0=3.25$ and $d\eta=0.25$) and
 exceeds an energy threshold.  The four momentum
sum of the pair is then used to compute the dijet mass ($M$), the
dijet transverse momentum ($k_T$) and the dijet longitudinal momentum
($p_z$).  Trigger efficiency corrections are then made, and the
yields are scaled by the measured integrated luminosity.  The dijet
cross sections for p+p collisions are shown in the right panel of Fig.~\ref{dijetcs}.

The combination of the forward inclusive jet cross section and the
forward dijet cross section is sensitive to parameters in PYTHIA such
as those that control parton showering and multi-parton interactions.
These parameters have been extensively tuned to fit jet cross
sections at the LHC \cite{Sk10}.  We find fair agreement between our forward jet
and dijet data and PYTHIA using the AUET2B tune developed by the ATLAS
collaboration \cite{ATLAS16}.  Dijet cross sections rapidly decrease
with $M$, $k_T$, and $p_z$.  There is no evidence for resonant
structures in the dijet $M$ interval shown in Fig.~\ref{dijetcs}.  The
jet threshold used in the figure is 30 GeV, but is arbitrary.  When increased to 45
GeV, it is found that AUET2B underpredicts the low-$M$ dijet yield.
The Perugia 0 tune \cite{Sk10} gives a good description of dijet cross
sections at both 30 and 45 GeV jet thresholds, but overpredicts the
inclusive jet cross section by more than a factor of 2.

\begin{figure}[!htbp]
  \centering
  \includegraphics[height=2.0in]{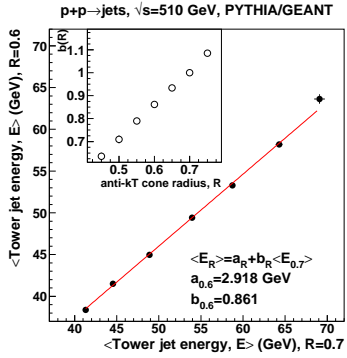}
  \caption{The jet energy scale is initially determined from the
    correlation of particle jet energy and tower jet energy.
    Correlations of tower jets with varying $R_{jet}$ versus tower
    jets with $R_{jet}=0.7$ establish that the jet-energy scale
    systematically changes with $R_{jet}$.  Results for $R_{jet}=0.6$
    are shown and the inset summarizes fits to correlations for $R_{jet}$ from
    0.4 to 0.7.  Energy compensation is used to restore the
    correlations with parton energy as $R_{jet}$ is decreased.}
  \label{jetscale}
\end{figure}

The $R_{jet}$ parameter in the anti-$k_T$ algorithm is frequently
varied as a means of discriminating jets from underlying event (UE)
contributions.  To preserve the meaning that a jet represents the
hadronization of a parton, it is necessary to adjust the jet energy
scale when $R_{jet}$ is varied.  Results from simulation are used
to relate tower jets to the parent parton, whose hadronization gives
rise to the jet.  Consequently, tower jets reconstructed with
different $R_{jet}$ are interrelated by their parent parton.  The jet
energy scale can be compensated as $R_{jet}$ is varied, as shown in
Fig.~\ref{jetscale}.  The slope of the energy
compensation is found to vary linearly with $R_{jet}$.  The UE
contributions are naively expected to be $\propto
\pi R_{jet}^2$.  It is likely
that specifics of the apparatus, such as granularity of the
calorimeter cells, affect this jet energy compensation.  The UE
contributions will also vary with $R_{jet}$ and may affect the
jet energy compensation.

\begin{figure}[!htbp]
  \begin{tabular}{ll}
    \includegraphics[width=1.85in]{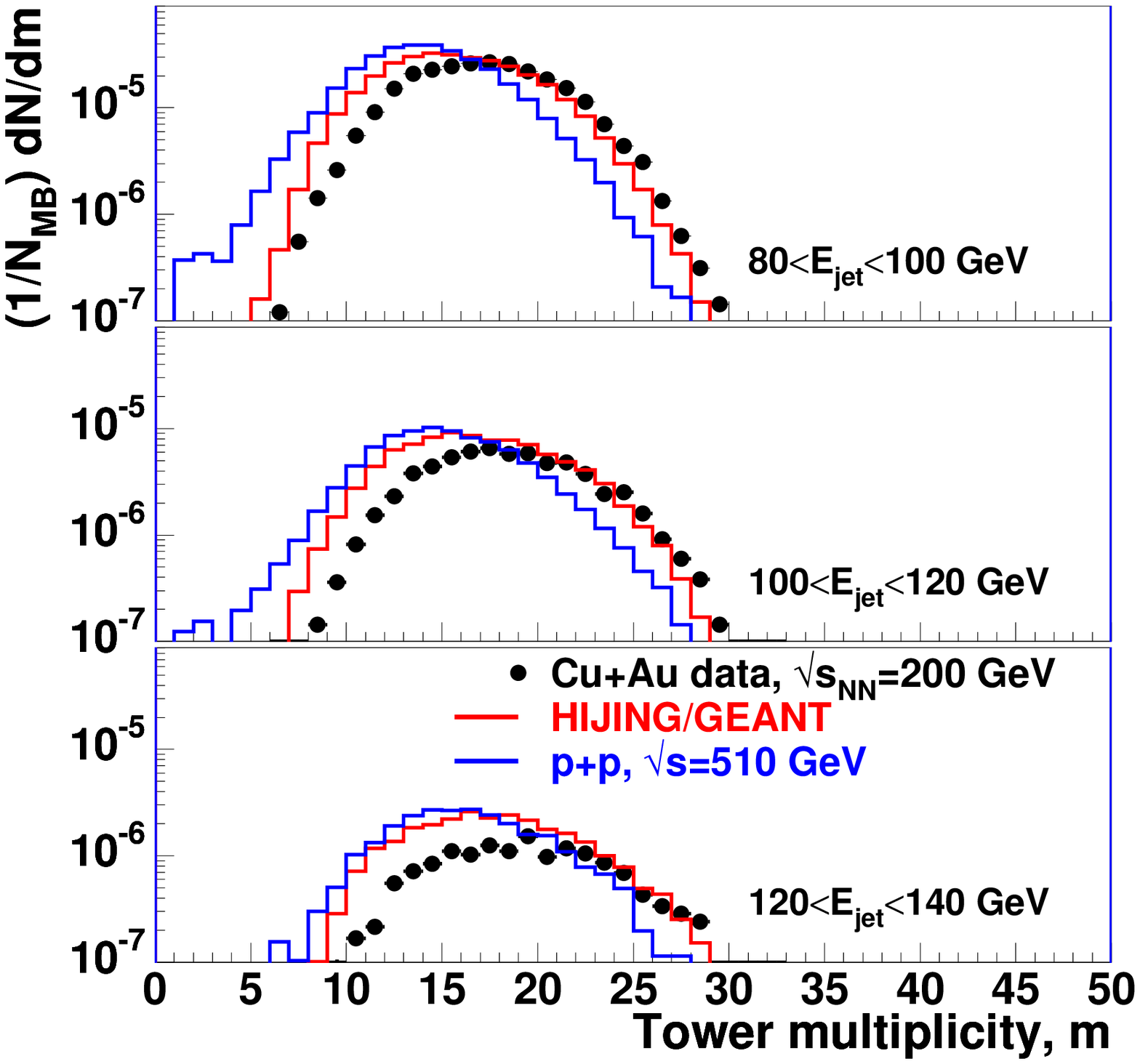} &
    \includegraphics[width=1.62in]{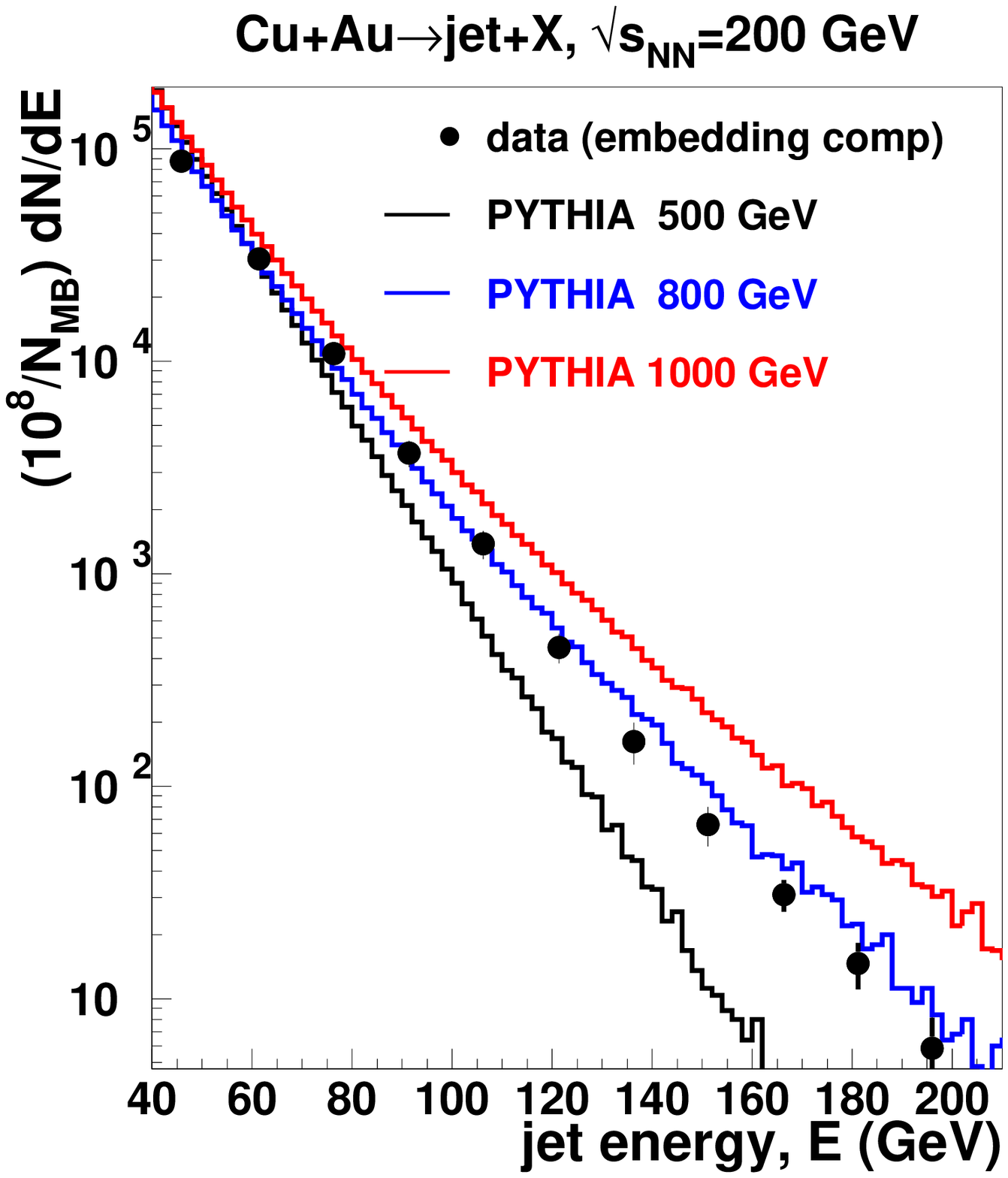}
  \end{tabular}
  \caption{(left) Tower multiplicity as a function of energy from
    patterns returned from the anti-kT jet finder, using $R_{jet}$=0.5 and
    energy compensation.  All distributions are scaled to represent
    the fraction of MB, with the restriction that $\Sigma Q_Y<8000$
    for Cu+Au.  Most features of Cu+Au data are represented by
    HIJING/GEANT simulation.  Bins with $E_{jet}>$100 GeV violate
    Feynman-x scaling for a single nucleon+nucleon collisions at
    $\sqrt{s}$=200 GeV.  The multiplicity distributions from Cu+Au jets are
    similar to those from pp jets produced in $\sqrt{s}$=510 GeV
    collisions. (right) Energy distributions for jets in Cu+Au collisions using $R_{jet}$=0.5
    and embedding compensation compared to jet spectra from p+p
    collisions at $\sqrt{s}>\sqrt{s_{NN}}$, normalized to the data at 50 GeV.
    The uncertainties plotted are the quadrature sum of statistical and systematic uncertainties.
    The systematic uncertainties are dominated by the jet energy scale, with
    contributions from run dependence.}
  \label{cuaujet}
\end{figure}

We now consider application of the anti-$k_T$ algorithm to Cu+Au
collision data.  The left panel of Fig.~\ref{cuaujet} shows that the
patterns returned by the jet finder for Cu+Au collisions have many
similarities to jets found in p+p collisions when $R_{jet}=0.5$ is
used with its energy compensation from Fig.~\ref{jetscale}.  A valid jet has $3.0<\eta_{jet}<3.5$.  For
the inclusive jet analysis we further impose $|\phi_{jet}-\pi|<0.70$ as
was done for the inclusive jet analysis in Fig.~\ref{dijetcs}.  This
region of the calorimeter had the response of a $4\times4$ submatrix
of cells split to allow dual ADC readout for the pulse-shape
discrimination tests.  This means that the range of energy deposition
sensitivity is doubled for these cells relative to the hardware gain used for p+p
data.  For data shown in Fig.~\ref{cuaujet}, there are essentially no
effects from ADC saturation, unlike at higher jet energies where ADC
saturation is clearly visible.  Features
observed in the Cu+Au data are also seen in GEANT simulations that use
the HIJING event generator \cite{WG91} to simulate Cu+Au collisions.
Analysis is restricted to semi-peripheral collisions by limiting the
total charge in the BBC annulus that faces the Au beam to $\Sigma
Q_Y<8000$.  This BBC annulus is $\sim$7 units of pseudorapidity
separated from the reconstructed jets.  A key feature is that the
forward jets in Cu+Au collisions extend well beyond the maximum energy
expected by $x_F$ scaling for single nucleon-nucleon collisions at
$\sqrt{s}=200$ GeV.  This is observed in both the data and the
HIJING/GEANT simulations.  We have confirmed that jet patterns from
towers selected randomly from events with similar vertex-$z$ and
$\Sigma Q_Y$ (mixed-tower analysis) have $2\times$ smaller average
tower multiplicity and have only 0.3\% of the yield in the left panel of Fig.~\ref{cuaujet}.

\begin{figure}[!htbp]
  \begin{tabular}{ll}
    \includegraphics[width=1.7in]{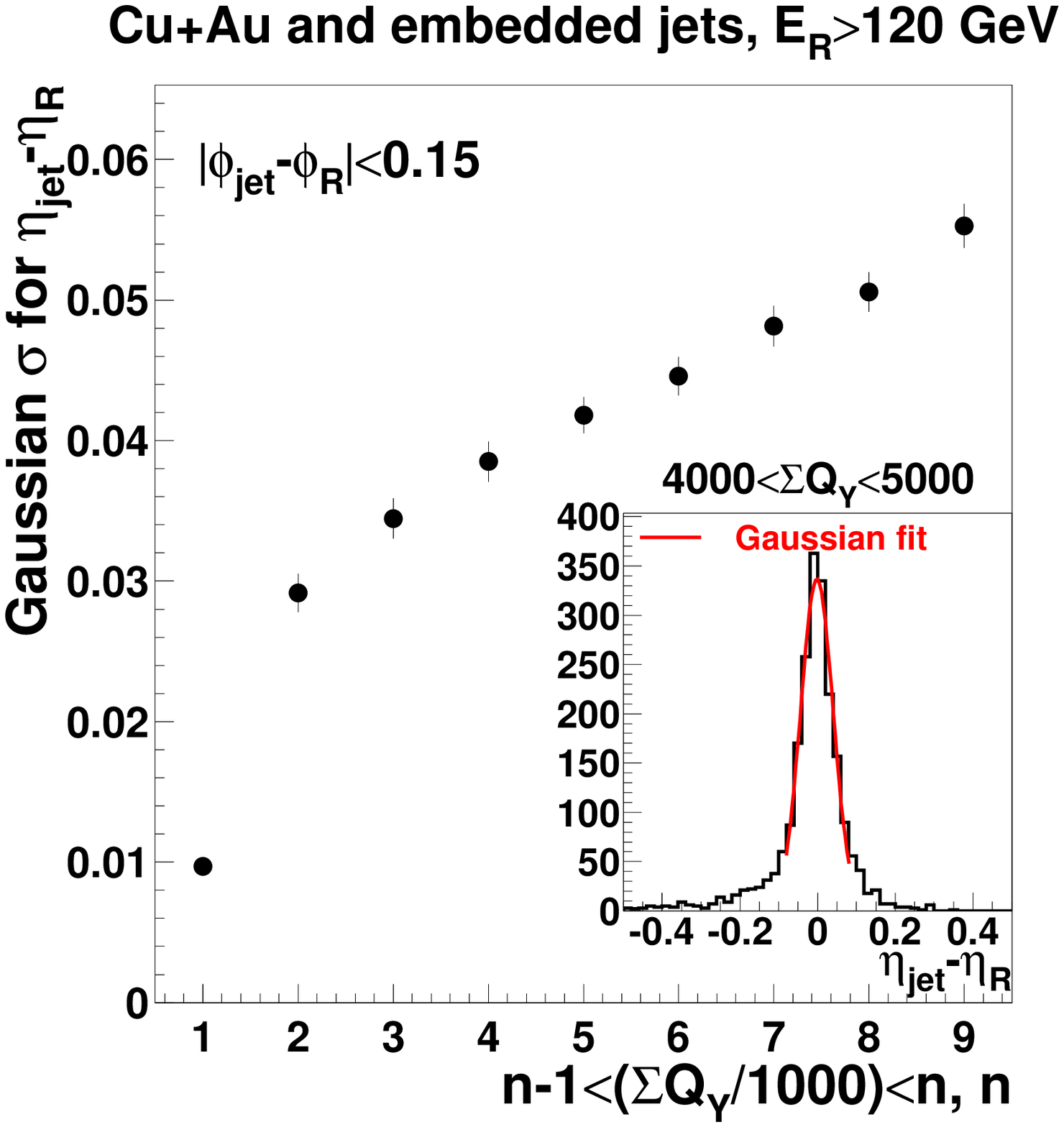} &
    \includegraphics[width=1.7in]{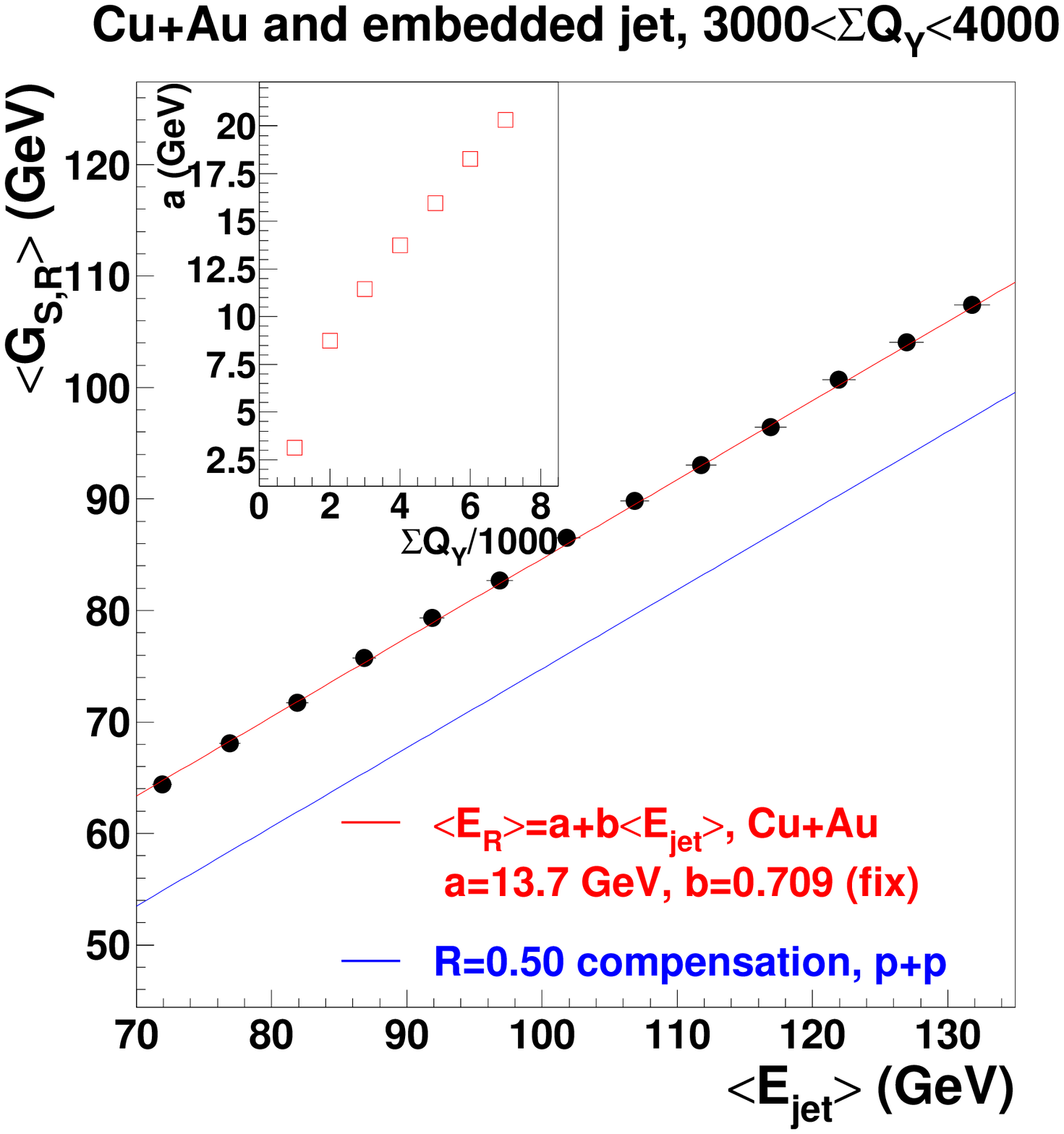}
  \end{tabular}
  \caption{ Results from embedding jets into Cu+Au data
    from a MB trigger, followed by reconstruction of the
    embedded event and comparison to the input jet.  (Left) Directional
    match resolution between reconstruction of embedded event and input jet in
    bins of $\Sigma Q_Y$, with the inset showing the distribution for
    one bin;
    (right) energy match between reconstruction of embedded event and
    input jet.  In general, jet reconstruction works, although
    underlying event contributions do impact the jet energy.}
  \label{embed}
\end{figure}

To quantify the impact of UE contributions in Cu+Au collisions on the
jet finding, jets from p+p collisions at $\sqrt{s}=510$ GeV are
embedded into Cu+Au MB data.  Normally, p+p reference data
is taken at $\sqrt{s}=\sqrt{s_{NN}}$ for the heavy-ion collision.  Jets from
p+p cannot exceed $x_F$ scaling limits, so embedding is done with
$\sqrt{s}=510$ GeV p+p collisions that result in the production of these high energy
jets.  The UE has little impact on the embedded jet
direction, as shown in the left panel of Fig.~\ref{embed}.  The
resolution smearing induced by Cu+Au UE is small
compared to the directional smearing between a parton and a jet in
p+p collisions.  The UE in Cu+Au collisions results in a
linear change to the energy of the embedded jet on average, as shown in the right
panel of Fig.~\ref{embed}.  This plots the average value of a skewed
Gaussian distribution ($\langle G_{S,R} \rangle$) that is fitted to the
reconstructed jet energy distribution, in a bin of embedded jet
energy.  Jets of these energies are minimally
impacted by Cu+Au UE, as evidenced from the slope term
of the energy compensation being the same in p+p and in Cu+Au.
The UE impacts the offset term in the energy compensation,
which linearly increases with the inverse of the impact parameter, as
observed by the linear dependence on $\Sigma Q_Y$ in the inset in the
right panel of Fig.~\ref{embed}.

We then apply the embedding compensation to the jets found in Cu+Au in
the right panel of Fig.~\ref{cuaujet}.  Jets extend beyond 200 GeV.
Checks were made that the slope of the jet energy spectrum is not
affected by energy dependence in the jet survival from embedding.
Further checks were made that the ADC for these high energy jets were
not saturated.  Results were verified using the Fastjet 3.3.2 sequential
recombination package \cite{FJ11}. Finally, results are similar between MB and
jet triggered data samples.  The highest jet energies violate Feynman
scaling by a factor of $>2$.  Such large violations were expected in
theoretical models \cite{Ar96} where QCD strings between color charges
released in the heavy-ion collision fuse, thereby increasing the
parton energy.  They called this a hadronic (QCD) accelerator.
A similar string fusion mechanism is in HIJING 2.10
\cite{WG91,De10} and AMPT \cite{Li05,Li01,Zh00}. 
HIJING 2.10 compares well to data in Fig.~\ref{cuaujet}. 
The EPOS model
\cite{Pi15} abruptly terminates particle production at $x_F=1$
assuming that nucleon collisions have $\sqrt{s} = \sqrt{s_{NN}}$.
Taking parton energy increase by string fusion literally, we compare 
Cu+Au jet data to p+p PYTHIA/GEANT
simulations with $\sqrt{s}>\sqrt{s_{NN}}$.  We show in
Fig.~\ref{cuaujet} that Cu+Au results at $\sqrt{s_{NN}}=200$ GeV 
are very similar to p+p collisions at $\sqrt{s}=800$ GeV.  The slope of
the jet energy distribution is sensitive to the equivalent $\sqrt{s}$
for p+p due to Feynman scaling limits on that distribution.  As for
our p+p measurements \cite{Bl15}, there is rapid
falloff of the yield as both the energy increases and the $p_T$
increases.  The $p_T$ dependence is well approximated by
$dN/dp_T \propto {\rm exp}(-p_T/p_{T0})$, with $p_{T0}\sim 1.30$ GeV/c.

\begin{figure}[!htbp]
  \begin{tabular}{ll}
    \includegraphics[width=2.05in]{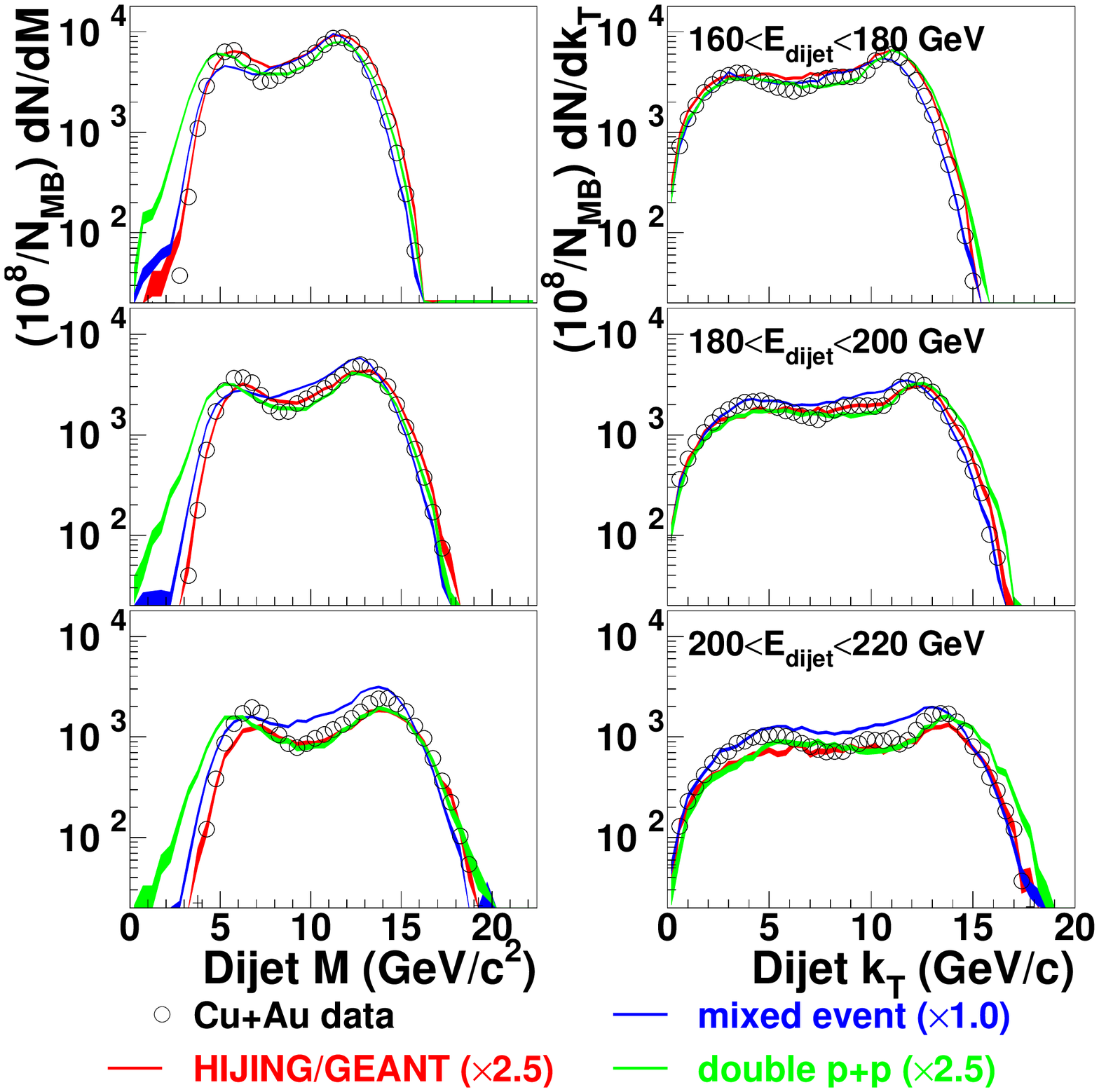}
    \includegraphics[width=1.40in]{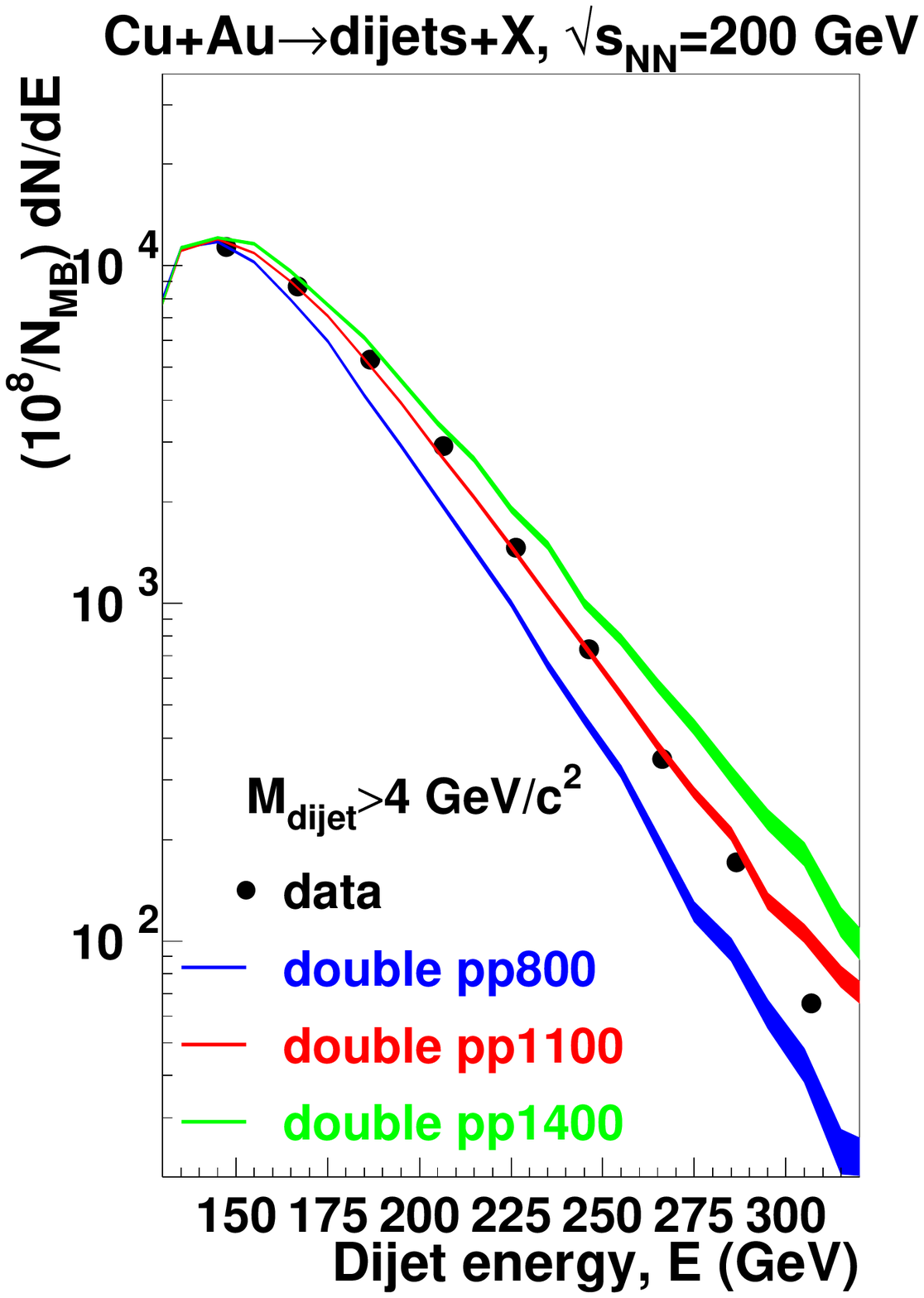}
  \end{tabular}
  \caption{(left) Uncorrected dijet distributions for Cu+Au data, in
    comparison to HIJING/GEANT simulation, results from a
    mixed-event analysis of the data, and double p+p interactions.  The left column of plots
    show the dijet mass distributions in bins of dijet energy and the
    right column of plots show the dijet momentum imbalance
    distributions in bins of dijet energy.  In general, HIJING/GEANT
    explains the data.  The agreement of mixed-event analyses with
    data and simulation suggest that here, jets are randomly
    produced. (right) Dijet dN/dE distribution for Cu+Au data in
    comparison to double p+p PYTHIA/GEANT events used to represent double
    parton scattering, normalized to data at 150 GeV. The uncertainties plotted are statistical.
    There is an estimated 10\% systematic uncertainty from calibrations and
    stability.} 
  \label{cuaudijet1}
\end{figure}

As for p+p, multiple jets are found in the acceptance in Cu+Au
collisions.  The four momenta of each jet are combined in the same manner
as for p+p.  Resulting $M$ and $k_T$ distributions are shown as a
function of dijet energy in Fig.~\ref{cuaudijet1}.  The dijet $M$
distributions are characterized by energy-dependent low-mass and high-mass peaks, which
arise because the acceptance is not perfectly annular.  The dijet energies
shown in this figure correspond to $1.6<x_F<2.2$, strongly suggesting
that multiple nucleons work together to give rise to the forward jet
pair since the $x_F$ value assumes single nucleon-nucleon collisions
at $\sqrt{s}=200$ GeV.  As for the inclusive jet results, HIJING/GEANT
simulations reproduce this behavior.  There is surprisingly good
agreement between simulation and data.  

Given the complexity of a relativistic heavy-ion collision,
we should expect for Cu+Au collisions that most jets are randomly
produced, meaning that jet pairs are not dynamically correlated for these energies.  To
test this, an event mixing algorithm is used on an ensemble of similar
events.  The ensemble requires that the collision vertex be common
within $\pm5$ cm and that $\Sigma Q_Y$ matches to within 200 ADC counts.
Towers for distinct jets are selected from random events, and then
added to create a mixed jet event.  The same reconstruction is applied
to the mixed-jet event as is applied to Cu+Au data and full simulation.
Nearly all features of both data and full simulation are explained by
mixed jet results in Fig.~\ref{cuaudijet1}.

The agreement between mixed events and Cu+Au dijet data prompts us to
compare the data to double parton scattering (DPS).  This is estimated by
analyzing pairs of p+p PYTHIA/GEANT events having vertex-$z$ within
$\pm0.5$ cm.  As for mixed events, the jets of such double events are
uncorrelated.  Unlike true DPS, UE contributions are
larger by including two p+p events.  Double p+p events, or uncorrelated DPS,
 can explain the bulk of the
Cu+Au dijet data taking $\sqrt{s}=1100$ GeV for the p+p collision
energy.  HIJING studies of the $p_z$ distributions of identified
particles require a distribution of p+p equivalent $\sqrt{s}$ for
$p_z>100$ GeV/c.  It is not unexpected that $\sqrt{s}$ in
Fig.~\ref{cuaudijet1} is larger than in Fig.~\ref{cuaujet} since the
dijet energies are larger than the jet energies.

\begin{figure}[!htbp]
  \centering
  \includegraphics[height=2.38in]{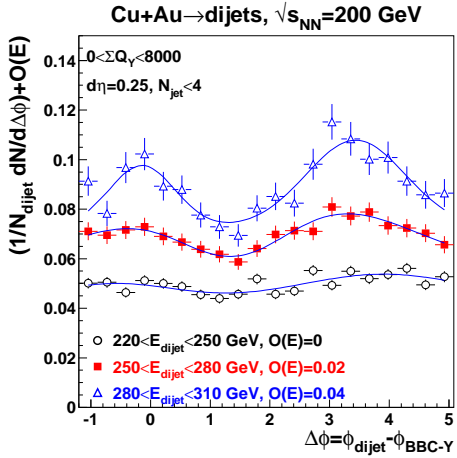}
  \caption{Azimuthal angle correlations between the dijet and particle
    multiplicity measured $\sim$7 units of pseudorapidity away.}
  \label{azim}
\end{figure}

Evidence for dynamical correlations of jet pairs appears as the dijet energy
increases.  Long-range rapidity correlations, possibly analogous to
those seen at the LHC \cite{CMS10}, become apparent at large dijet energy, as seen
in Fig.~\ref{azim}.  These correlations have been explained in a string fusion
\cite{Br15} or flux tube picture \cite{Bj13}.
The dijet has a transverse momentum ($k_T$)
directed at $\phi_{dijet}$.  The particle multiplicity observed in the annular tiles of the
BBC facing the Au beam has a charge-weighted average orientation
$\phi_{BBC-Y}$.  The angle difference
$\Delta\phi=\phi_{dijet}-\phi_{BBC-Y}$ is shown in Fig.~\ref{azim}.
As the dijet energy increases, peaks near $\Delta\phi=0$ and
$\Delta\phi=\pi$ become evident.  The latter peak can be from momentum
conservation, but the former peak is not expected except at small
$\Delta\eta$.  The pseudorapidity separation of the dijet from the
measured particle multiplicity is $\Delta\eta\sim 7$.  An additional
condition limiting the number of good jets in the event to $<4$ is
imposed so as to reduce combinatoric backgrounds.

\begin{figure}[!htbp]
  \centering
  \includegraphics[height=3.38in]{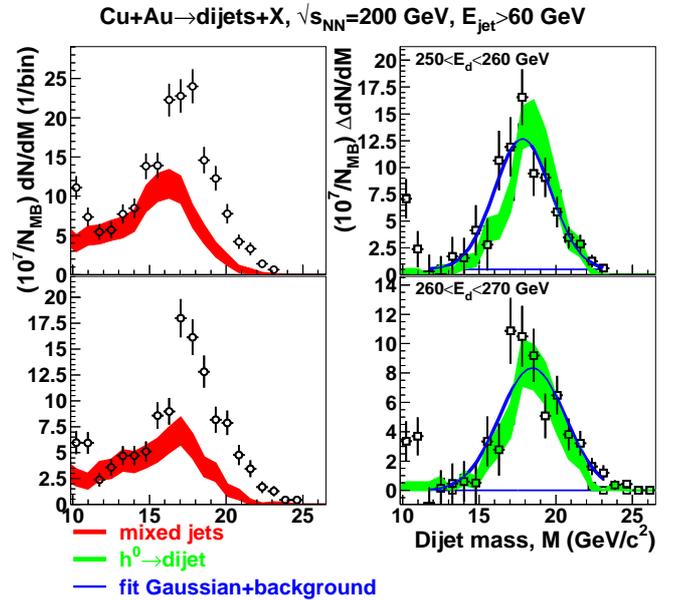}
  \caption{Dijet mass compared to a mixed-event analysis in the left
    column.  The right column forms the difference between data and
    mixed events, and compares that difference to a simulation of the
    production of a resonance that decays to jet pairs.  All Cu+Au
    distributions have vertical axes scaled as $10^7/N_{MB}$.}
  \label{cuaudijet2}
\end{figure}

To investigate these correlations further we  look at
dijet mass at large dijet energies in Fig.~\ref{cuaudijet2}.
As in Fig.~\ref{cuaudijet1}, there are
energy-dependent mass peaks in mixed events.
The difference is formed between the data and
mixed-jet events, and is shown in the right column of
Fig.~\ref{cuaudijet2}.  This difference can be fit with a Gaussian
distribution with small remnant background contributions.  
Peaks are apparent
at $M=17.83\pm0.20$ GeV/c$^2$ in the $250<E<260$
GeV dijet energy bin and at $M=18.47\pm 0.22$ GeV/c$^2$ in the
$260<E<270$ GeV dijet energy bin. 
The
statistical significance of the peaks are 9.0 standard deviations in
the dijet energy bin $250<E<260$ GeV and 8.4 standard deviations in
the dijet energy bin $260<E<270$ GeV.  Dijet mass background can be
described by mixed jet events, by random jet pairs, and by HIJING/GEANT,
with decreasing importance as the dijet
energy increases.  All of these methods yield essentially
the same results, with the peak centroids varying little
from the means in Fig.~\ref{cuaudijet2}.
The dijet peak is evident down
to dijet energies of $\sim$240 GeV.  At lower dijet energies the mass
distribution is predominantly explained by event mixing until near
the $\chi_b$ region, where contributions from heavy hadrons are evident.
 Dijet energies higher than $270$ GeV are increasingly affected by ADC
saturation.  Given that there is little yield above $M\sim 12$
GeV/$c^2$ except for the observed peak, local and global statistical
significance are the same.  Combining statistical uncertainties for
the two bins results in $M=18.12\pm0.15$ GeV/$c^2$.

Systematic studies of dijet mass were conducted.  The mass peak is
present for inclusive pairing of all good jets in the acceptance,
including events where the energy sum in the perimeter of
cells closest to the beam, $E_{P1}$, exceeds 350 GeV.
Events with $E_{P1}>350$
GeV are excluded in Fig.~\ref{cuaudijet2} as in Fig.~\ref{cuaudijet1}.
This near-beam energy sum is not strongly
correlated with particle multiplicity in the BBC.  Also imposed in
Fig.~\ref{cuaudijet2} is a requirement
that the jet patterns do not have saturated ADC values.  The mass peak
is present with or without this requirement.
Another event selection that reduces
combinatoric backgrounds is to limit analysis to events that have the
number of good jets less than 4.  Systematic studies included
variation of the portion of the vertex-$z$ distribution chosen for the
analysis, with the nominal selection $|z_v|<75$ cm varied to
$|z_v+30|<75$ cm.  
A systematic uncertainty of $\pm0.13$ GeV/$c^2$
is estimated from the root-mean square of values from
varying the event selection and the portion of the vertex-$z$
distribution used in the analysis.
The dijet mass peak centroid is
stable for different event selections and varying
jet background estimates, and is found to be $18.12\pm0.15~(stat)\pm0.6~(sys)$ GeV/c$^2$.

Further systematic studies were conducted to establish if instrumental
effects were responsible for the dijet mass peak.  The cell with the
largest energy deposition of those found for the dijet is distributed nearly uniformly
over the azimuth.  The dijet mass peak is not due to a small number of
calorimeter cells. Given that energy depositions far exceed $x_F$
scaling limits, it was also examined if saturation of the electronics
was responsible for the mass peak.  The peak is present for dijet
patterns that do not saturate the electronics.  It is also ruled out
that special conditions of the colliding beams are responsible, by
finding that yields of the dijet mass peak are relatively constant for
the data taking period.  Acceptance requirements and collision vertex
requirements were also varied, with minimal effect on the mass peak.
We conclude that the dijet mass peak in Fig.~\ref{cuaudijet2} is not
an effect of the instrumentation.

The question then becomes whether it is plausible that a new particle
can give rise to a peak in the forward dijet mass.  To address this, we
used matrix elements for $p+p\rightarrow h^0+X$ available in PYTHIA,
at $\sqrt{s}=1200$ GeV chosen to result in subtantial production probability for
$h^0$ in the energy range from $E>250$ GeV.  The resonance mass was
adjusted in the simulation to be $M=18.2$ GeV/$c^2$ and the full width
of the resonance was left at the default of 20 keV.  Decays of $h^0$ were limited to jet
pairs.  GEANT simulations were run on the simulated events, and
reconstructions of these events were then done.  The resulting dijet
mass distribution is overlayed with background subtracted data in the
right column of Fig.~\ref{cuaudijet2}.  Production of a resonance that
decays to two jets describes the centroid and width of the Cu+Au dijet
data.  It was further confirmed that the jets reconstructed matched
the directions and energies of the parton decay daughters of the
resonance.  Since the input resonance width is small, the dijet width
is limited by the resolution of jet finding and detector effects.
Model studies of resonance production show that the opening angle
between the reconstructed jets does not match the opening angle
between the resonance daughters, resulting in energy dependence to the
reconstructed mass.  This is attributed to finite acceptance effects,
which model studies show are small at the energies in Fig.~\ref{cuaudijet2}.

\begin{figure}[!htbp]
  \begin{centering}
    \begin{tabular}{ll}
      \includegraphics[width=1.5in]{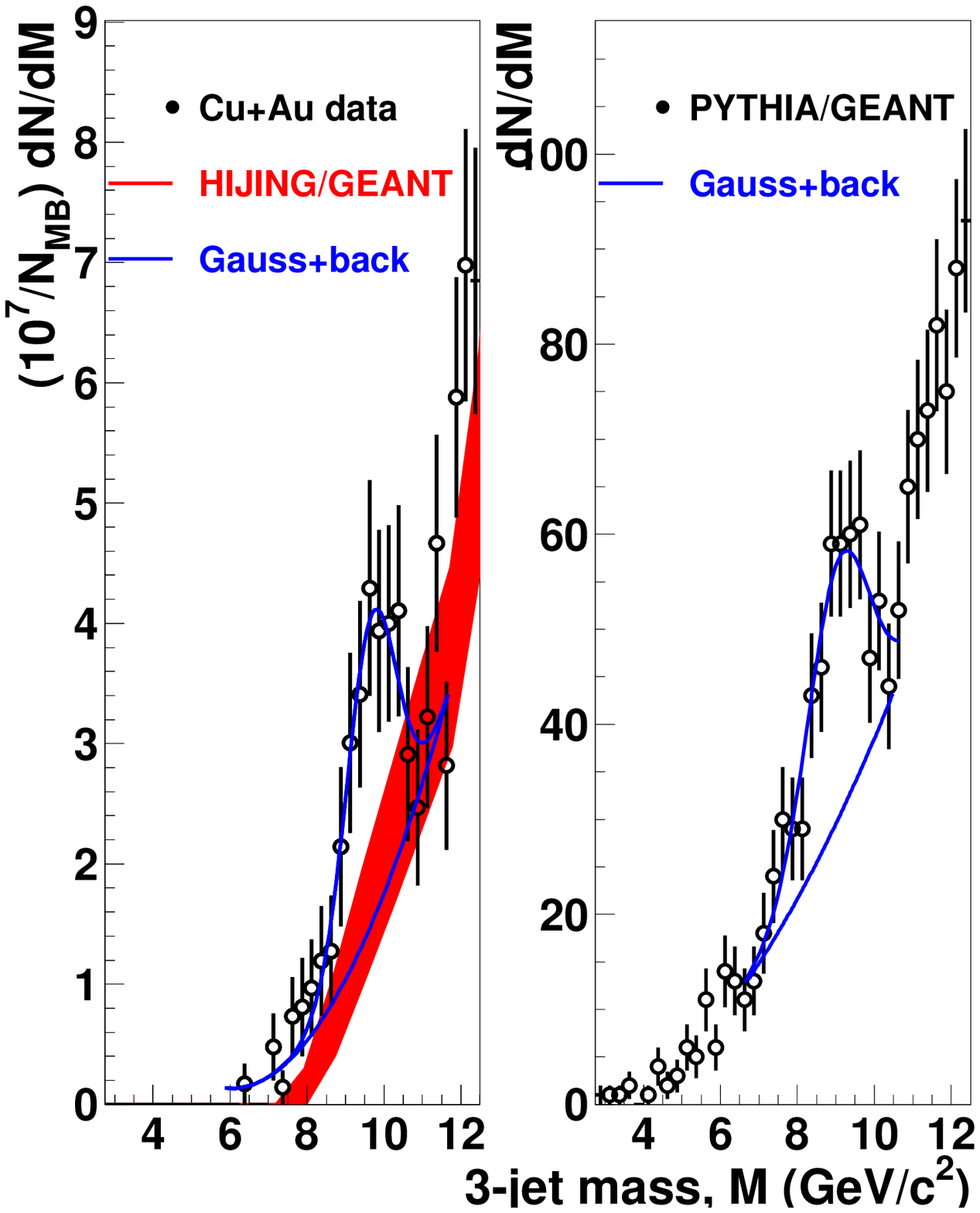}
      \includegraphics[width=2.0in]{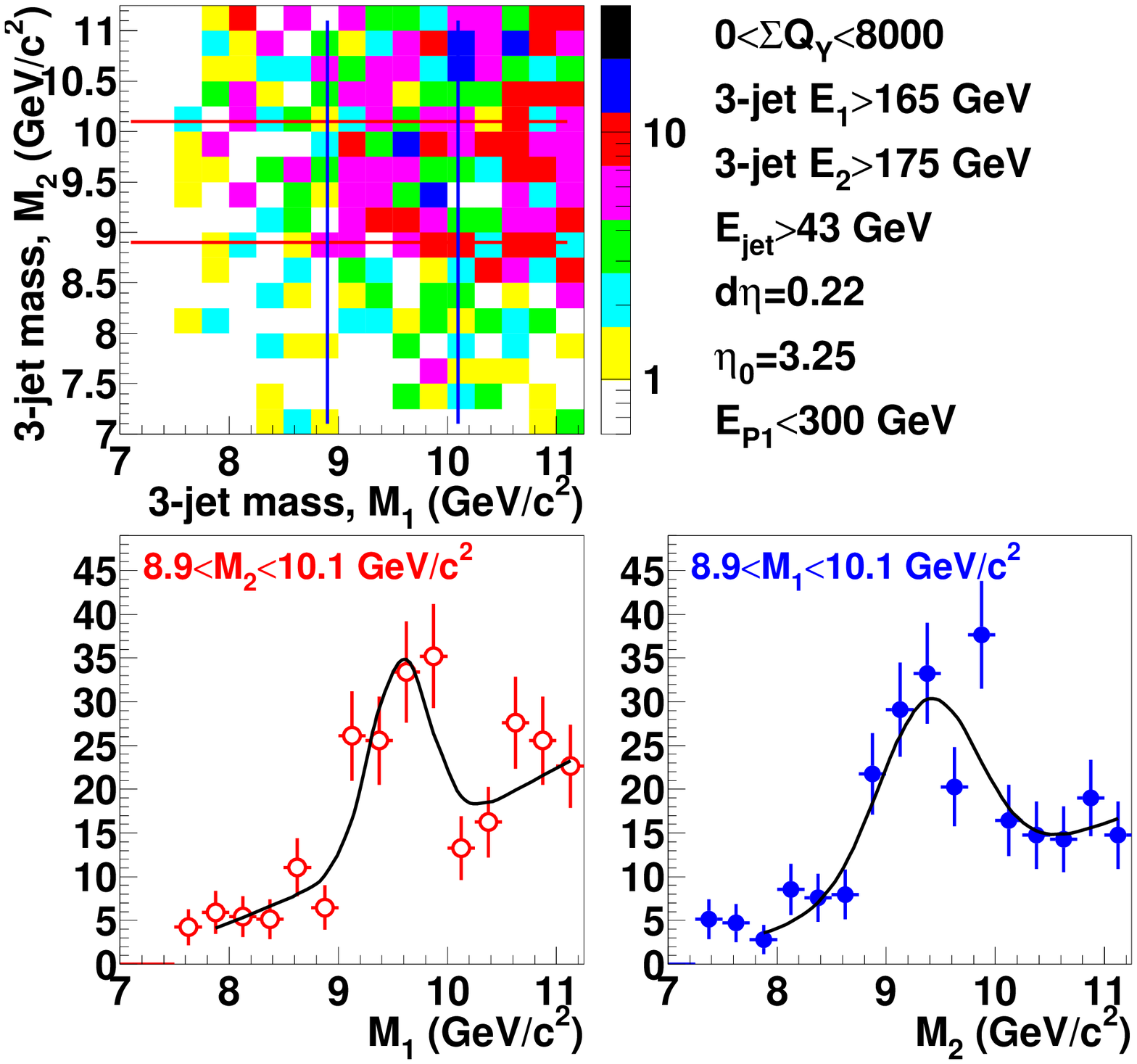}
    \end{tabular}
  \end{centering}
  \caption{Evidence for $\Upsilon$(1S) via its decay to three jets.
    (left pair) Inclusive forward production from Cu+Au collisions
    overlayed with HIJING/GEANT simulation.  A $5.2\sigma$ peak is
    observed in the data.  Comparison is to PYTHIA/GEANT p+p
    simulations at $\sqrt{s}=1200$ GeV, using the Perugia 0 tune.
    (right) $\sim$5$\sigma$ evidence for forward pair $\Upsilon$(1S)
    production.  All Cu+Au distributions have vertical axes scaled as
    $10^7/N_{MB}$.}
  \label{upsilon}
\end{figure}

Checks of the jet energy scale with embedding compensation were made
by extending the analysis from dijets to combinations of larger number of jets.  The
left panel of Fig.~\ref{upsilon} shows results for the inclusive
3-jet mass, where valid jets are within the acceptance with energy
$>60$ GeV.  HIJING/GEANT simulations are also shown, and describe the
increasing background as 3-jet mass increases.  A peak in Cu+Au data
is evident with statistical significance of 5.4 standard deviations.
Simulations of $\Upsilon$(1S) production with either PYTHIA/GEANT or
PYONIA/GEANT have been studied, and also result in a 3-jet mass peak.
Consequently, we attribute the peak in 3-jet mass from Cu+Au
collisions to $\Upsilon$(1S)$\rightarrow 3g$, studied in $e^+e^-$
collisions \cite{SM10}.  Production of $\Upsilon$(1S) in p+p
collisions at the energy shown in Fig.~\ref{upsilon} is not possible
for $\sqrt{s}=\sqrt{s_{NN}}$ because it is beyond the kinematic limit
for the rapidity acceptance of the data.  However, via parton energy increase
by string fusion in Cu+Au collisions, $\Upsilon$(1S) production
is feasible for $\sqrt{s}=1100$ GeV p+p collisions, which is the
p+p equivalent collision energy deduced from Fig.~\ref{cuaudijet1}.  Yield
determinations from 3-jet reconstructions are uncertain because of
sensitivity to UE contributions.  There is no heavy-ion model that
includes string fusion and proper treatment of heavy-quark or
$\Upsilon$ production.  Consequently, our simulation studies are
restricted to PYTHIA.  There is strong tune dependence to whether
these simulations produce a 3-jet mass peak, and there is evidence that
UE contributions in Cu+Au for Feynman scaling violations is smaller
than in high-energy p+p collisions.  The known mass \cite{PDG} of
$\Upsilon$(1S) constrains the jet energy scale to $\pm2$\% using a
method described in \cite{Bl15}.

The right panel of Fig.~\ref{upsilon} shows $\sim$5 standard deviation
evidence for double $\Upsilon$(1S) production, where each $\Upsilon$
is reconstructed from 3-jet combinations.  Given that this analysis
has little combinatoric background, the jet energy threshold is lowered
from 60 GeV used for the inclusive 3-jet analysis to 43 GeV, used
to search for double $\Upsilon$ production.  The lower threshold increases
the efficiency for $\Upsilon\rightarrow 3g$, given the energy distribution
of the gluons \cite{KW78}.  Although the acceptance for
this is small, such production is expected from DPS, as for jet pairs
in Cu+Au collisions.  Double $\Upsilon$(1S) reconstruction is evident
in analysis of double PYONIA/GEANT events.  There is no 
UE in those simulations by construction. 

Based on the dijet mass peak position (Fig.~\ref{cuaudijet2}) and
evidence for single and double $\Upsilon$(1S) production
(Fig.~\ref{upsilon}) , the most likely candidate for the dijet mass peak is an
all-$b$ tetraquark, $X_b$, a configuration of
$b\overline{b}b\overline{b}$.  There are many recent theoretical
calculations of the mass of this object.  Karliner, Rosner, and
Nussiov \cite{KRN17} (KRN) make estimates of the mass of $X_{b}$ based
on systematics of meson and baryon masses.  They estimate
$M_X=18.826\pm0.025$ GeV/$c^2$.  They also estimate a production
cross section of 1 pb for p+p interactions at the LHC ($\sqrt{s}=13$
TeV), based in part on a report of double $\Upsilon(1S)$ production
\cite{CMS16} in p+p collisions at $\sqrt{s}=8$ TeV.  Further, KRN
state that if $X_{b}$ is significantly lighter than their estimate,
then decays to both $\gamma\gamma$ and $gg$ become favorable. 
 Analogous to KRN, the
Cu+Au data here has significant $\Upsilon$(1S) production as seen
through its 3-jet decay. In addition, significant double $\Upsilon$(1S)
production is observed in the Cu+Au data, most likely via a DPS
mechanism.  These observations are compatible with the interpretation
that the dijet mass peak in Cu+Au is due to an all-$b$ tetraquark.
The exploratory nature of these measurements preclude accurate
determination of yields.

Many other authors have also considered the existence of an all-b tetraquark.
Bai, Lu, and Osborne (BLO) \cite{BLO16} estimate $M_X=18.69\pm0.03$ GeV for the
ground state of the all-$b$ tetraquark configuration  \cite{QR79}.
Our measured peak position $\mu=18.12\pm0.15 (stat)$
GeV/$c^2$ is significantly smaller than both KRN and BLO.  It is possible that
double $b\overline{b}$ annihilation, resulting in $X_b\rightarrow gg$
becomes the preferred decay mode as the all-$b$ tetraquark mass becomes
smaller.  Initial searches for an all-$b$ tetraquark in lattice QCD have found no
evidence for its production \cite{HED18}.
Richard, Valcarce, and Vijande \cite{RVV17}, conclude that with
a rigorous treatment of the four-body problem, $X_b$ is unbound.  Wu,
{\it et al.} \cite{Wu17} find the ground state configuration for $X_b$
to have a mass of 18.46 GeV/$c^2$. Wang \cite{Wa17}
uses the method of QCD sum rules to predict the ground state mass of
$X_b$ as $M=18.84\pm0.09$ GeV.  These references
\cite{Wa18, Es18, Ch18, An17} also discuss theoretical aspects of
all-$b$ tetraquarks.

Experimental results from the LHC are also now becoming available.
LHCb searches for an all-$b$ tetraquark via its decay to
$\Upsilon$(1S)$+\Upsilon^*\rightarrow \mu^+\mu^-\mu^+\mu^-$
\cite{LHCb18}.  Searches were made in p+p data samples at $\sqrt{s}=7,
8,$ and $13$ TeV.  The muons were detected in the pseudrapidity range
from $2<\eta<5$.  They set limits on production of an all-$b$
tetraquark.  In addition, CMS has conducted a search in p+p data
samples at $\sqrt{s}=7$ and 8 TeV looking at
$\Upsilon$(1S)$+\Upsilon^*$ detected by $\mu^+\mu^-\mu^+\mu^-$ and
$\mu^+\mu^-e^+e^-$ at midrapidity.  A preliminary report \cite{Du18}
finds an all-$b$ tetraquark candidate at $M=18.4\pm 0.1 (stat) \pm 0.2
(sys)$ GeV/$c^2$ with 3.6 standard deviation significance.  This
preliminary result has a mass peak in good agreement with the
dijet signal we observe in Cu+Au collisions at $\sqrt{s_{NN}}=200$
GeV.

In conclusion, we have observed forward jets produced in Cu+Au
collisions at $\sqrt{s_{NN}}=200$ GeV.  We know these are real jets from
embedding studies, from mixed-tower analyses that do not match the jet
data, and from observation of $\Upsilon$(1S) through its 3-jet decay.
We understood and verified jet energy scale from reconstruction of $\Upsilon$(1S)$\rightarrow 3g$.
Jet and dijet results were verified using 2 independent analyses, one based on
the anti-kT algorithm and the other based on the Fastjet 3.3.2 package.
The jets produced in Cu+Au collisions exceed Feynman scaling limits by
a factor of 2, assuming that the equivalent p+p
$\sqrt{s}=\sqrt{s_{NN}}$, as is commonly done.  Parton energy increase
from string fusion (QCD accelerator) can explain these scaling violations.  The
jet data can be explained by p+p simulations at $\sqrt{s}=800$ GeV.
Jet pairs produced in Cu+Au collisions are also observed.  They can be
explained as double parton scattering, with a parton flux matching p+p
collisions at $\sqrt{s}=1100$ GeV.  Further evidence of parton energy
increase is obtained from our observation of $\Upsilon$(1S),
through its 3-jet decay.  Double $\Upsilon$(1S) production is also
observed.  We observe long-range rapidity correlations for dijets
with energies greater than 250 GeV. Finally, dijet mass in Cu+Au
collisions shows evidence of a 
signficant peak at $M=18.12\pm0.15 (stat)\pm0.6(sys)$ GeV/$c^2$.  Our results are
compatible with the first observation of an all-$b$ tetraquark.

As an outlook, the parton energy increase mechanism evident in the forward
direction opens up prospects for searches for particles that either
probe physics beyond the standard model ({\it e.g.} axion-like
particles \cite{BNT17}) or are relevant to cosmology ({\it e.g.} dark
matter \cite{Ba17}).  Appropriate forward instrumentation that can
handle the high energies from string fusion can probe for such
particles that decay to jet pairs, in a dijet mass range from $5 \le M
\le 50$ GeV/$c^2$.  Di-photon decays could also be explored.
The large amount of color charges released in a relativistic heavy-ion
collision are expected to be fertile territory for new particle
searches.

We thank the RHIC Operations Group at BNL.  This work was supported in part
by the Office of NP within the U.S. DOE Office of Science, the Ministry of Ed.
and Sci. of the Russian Federation, and the Ministry of Sci., Ed. and Sports
of the Rep. of Croatia, and IKERBASQUE and the UPV/EHU.

%\end{linenumbers}

\end{document}